\documentstyle[12pt]{article}
 \advance\oddsidemargin by -1in
\advance\evensidemargin
 by -1in
\advance\topmargin by -0.0in
\makeindex

\newcommand\beq{\begin{equation}}
 \newcommand\eeq{\end{equation}}
\newcommand\beqn{\begin{eqnarray}}
 \newcommand\eeqn{\end{eqnarray}}
\newcommand{\doublespace} {
 \renewcommand{\baselinestretch} {1.6}
\large\normalsize}

\begin{document}
\vspace*{4cm}

 \centerline{\Large \bf Deep-Inelastic Electroproduction of Neutrons}
\medskip
 \centerline{\Large \bf in the Proton Fragmentation Region}

\vspace{.5cm}
\begin{center}
{\large Boris~Kopeliovich\footnote{On leave from Joint Institute for
 Nuclear Research, Laboratory of Nuclear Problems,
\newline Dubna, 141980 Moscow Region, Russia.  E-mail:
 bzk@dxnhd1.mpi-hd.mpg.de}, Bogdan Povh}

\vspace{0.3cm}

 {\sl Max-Planck Institut f\"ur Kernphysik, Postfach 103980,
\newline 69029 Heidelberg, Germany}\\

\vspace{0.3cm}

and

\vspace{0.3cm}

{\large Irina Potashnikova}

\vspace{0.3cm}

{\sl Joint Institute for
 Nuclear Research, Laboratory of Nuclear Problems,
\newline Dubna, 141980 Moscow Region, Russia}

\end{center}

\vspace{1cm}
\begin{abstract}

Experiments at HERA looking for deep-inelastic
electroproduction of neutrons in the proton fragmentation
region are in progress.
They are aimed to measure the pion
structure function at small Bjorken $x$. The important condition for
such a study is to establish under what kinematical conditions 
the dominance of the pion-pole graph in the process is guaranteed.
We analyse other sources of the 
leading neutron, in order to
figure out the kinematical region where the one-pion exchange
dominates.

\end{abstract}

\newpage

\doublespace
\section{Introduction}

\begin{figure}[tbh]
\includegraphics{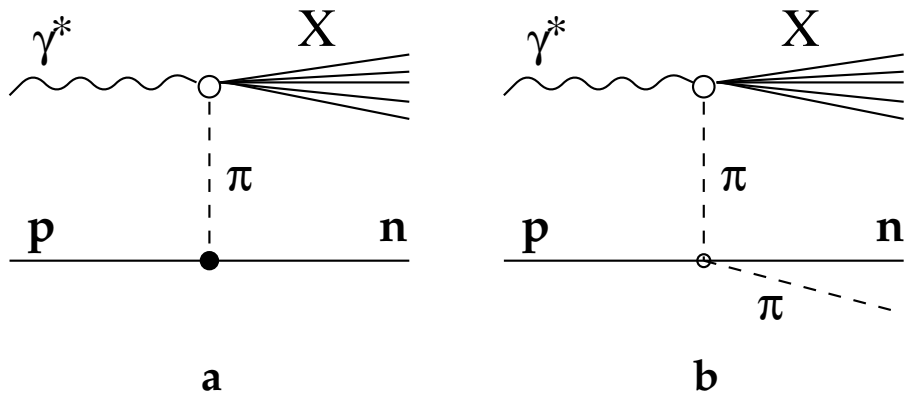}
\begin{center}
\vspace{5cm}
\parbox{13cm}
{\caption[Delta]
{One-pion-exchange diagrams contributing to the deep-inelastic
scattering $ep \to e'nX$ 
without ({\bf a}) and with ({\bf b})
production of an additional pion in the proton fragmentation region.}
\label{fig1}}
\end{center}
\end{figure}

Deep-inelastic lepton scattering off a proton at small Bjorken $x$
probes the distribution of sea quarks and antiquarks in the
proton. The cloud of virtual pions surrounding the nucleon core of the
proton is known to be an important source of the sea quarks \cite{abk}.
Thus, one may hope to be able to extract information about the pion
structure function from such measurements, relying upon the dominance 
of the one-pion-exchange diagram shown in fig.~\ref{fig1}a 
\cite{sull,desy,juelich}.

One has to provide, however, convincing evidence that the
sea quarks probed by the virtual photons do really belong to
the pion. There are many other sources of sea quarks 
the contribution of which has to be evaluated reliably.
Moreover, even if one is sure that the probed 
sea quark originates from a
pion, in order to extract the pion structure function from the 
data, one relies upon the diagram in fig.~\ref{fig1}a. Another
one depicted in fig.~\ref{fig1}b (as well as others with additional 
particle production in the pion-proton interaction vertex) should be
considered as an unwanted background, because they result in a different
relation between the deep-inelastic scattering cross section and 
the pion structure functions.

In the present paper we study the background mechanisms of
the leading neutron production in deep-inelastic scattering, which
may distort the determination of the pion structure function.
We conclude that there is a possibility to suppress the backgrounds 
if the neutron is detected within $z$-interval $0.7-0.9$ and at 
$|t| < 0.2-0.3\ GeV^2$.

Although the recommendations of the 
previous analysis \cite{sull}-\cite{juelich}
are not very far from our conclusion, our results
are quite different at some essential points concerning the pion pole 
contribution, as well as that 
for the background. Furthermore, as the diagram in fig.~\ref{fig1}b
cannot be evaluated using the light-cone representation for the nucleon
wave function, we therefore apply the standard Feynman diagram
technique.

\section{The signatures of the pion}

What is specific about the pion, which may help to single out its
contribution as compared to other
mesons surrounding the proton? Of course it is the smallness of the pion mass.
It allows the pion to be spread around the proton at much longer distances
than other mesons. This fact manifests itself as a narrow 
transverse-momentum distribution, $\langle p_T^2 \rangle \sim m_{\pi}^2$.
The same argument is applicable to the diagram in fig.~\ref{fig1}b, but
the narrow peak in the $p_T$ distribution of the neutrons from the process
 in fig.~\ref{fig1}a will be smeared out by the emission of
the additional pion, which depends on the effective mass of the final
$\pi n$ system.

Thus, an observation of a narrow peak in the $p_t$-distribution of
the produced neutrons would serve as a solid evidence of the pion-exchange
dominance.

The above arguments, however, fail if the pion is far
off the mass shell. This depends on the energy $\nu$ of the virtual photon and
the effective mass $M$ of the produced jet.
Indeed, if the pion in the rest frame of the proton has a ''mass'' $\sqrt{|t|}$,
where $t<0$ is the pion four-momentum squares, the photon cannot
produce a jet, with an effective mass larger than $M^2 \approx 
2\nu \sqrt{|t|}$.
Thus, the production of a heavy mass jet, $(M^2/2\nu)^2 \gg m_{\pi}^2$
needs a highly virtual pion target, $|t| \gg m_{\pi}^2$, and 
the width of $p_T^2$ distribution of the produced neutrons
is determined by $|t|$ rather than by $m_{\pi}^2$.
This means that such a far-off-mass-shell pion does not propagate
far away from the proton, but only for a short distance of about $1/\sqrt{|t|}$.
This area is overpopulated by sea quarks having other than a pion origin,
i.e. the pion loses its signature.

\medskip

\begin{figure}[tbh]
\includegraphics{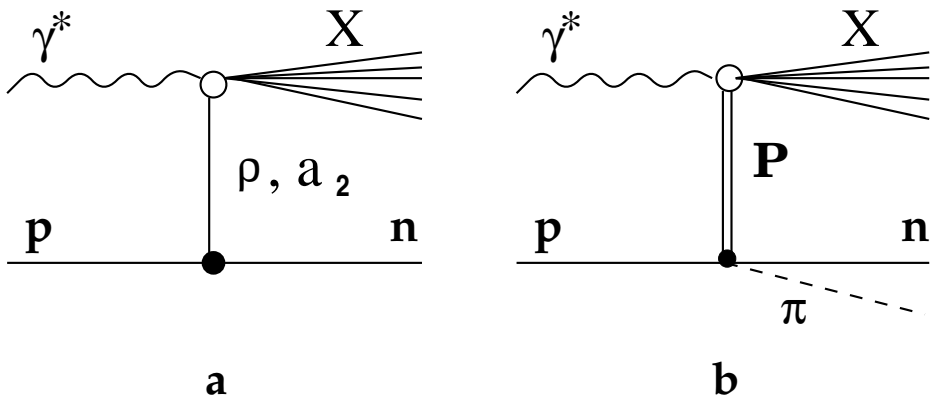}
\begin{center}
\vspace{5cm}
\parbox{13cm}
{\caption[Delta]
{The $\rho$- and $a_2$-Reggeon ({\bf a}) and the Pomeron ({\bf b}) 
exchange contributions to the deep-inelastic
scattering $ep \to e'nX$.}
\label{fig2}}
\end{center}
\end{figure}

In order to make $|t|$ sufficiently small and
single out the pion contribution one may select the events with a small value of 
$(M^2/2\nu)^2 \leq m_{\pi}^2$. However, in this case
one faces another problem, which is related to the spinelessness of the
pion. It is true that the pion contribution is enhanced at large 
transverse distances from the proton, what corresponds to small $\langle
p_T^2 \rangle$ of the recoil neutron. 
However, the rapidity gap covered by the virtual pion
becomes large in this case, $\Delta y \approx \ln(1-z)$, where $1-z \approx
M^2/2m_N\nu \leq m_{\pi}/m_N$. We know that the energy dependence of the scattering
amplitude corresponding to a particle exchange is controlled by the spin of the particle.
In the Regge theory the Reggeon intercept plays the role of the spin. It is about the
same as the spin for the pion, $\alpha_{\pi}(0)\approx 0$; 
is higher for leading Reggeons $\rho$ and
$a_2$, $\alpha_R(0)\approx 1/2$; and it has the highest value for the Pomeron,
$\alpha_P >1$. Therefore, the probability for a 
pion to cover a large rapidity gap
is suppressed by the factor $\exp(-2\Delta y)$, 
which is smaller than the corresponding factor  $\exp(-\Delta y)$ for the leading
Reggeons $\rho$ and $a_2$ (see below), or for
the Pomeron, which is rapidity-independent, 
or even grows with $\Delta y$. 
In such circumstances one should 
be very cautious about the competing mechanisms
of generation of the sea quarks.
Examples of diagrams corresponding 
to the contribution of these mechanisms to the neutron production are 
shown in fig.~\ref{fig2}. 

We conclude that any attempts to single out the pion-pole contribution
relying upon the signatures of the pion, the smallness of its mass and
spinelessness, face kinematical restrictions which in some sense exclude
each other. In such circumstances one should look for
a compromise, i.e. a kinematical region where the pion pole
nevertheless dominates. This is the goal of the present paper.

\section{The pion-pole contribution to $ep \to e'nX$ (fig. 1a)}

First of all, we should calculate the diagram shown in fig.~\ref{fig1}a,
which is to be dominant in order to extract the pion structure
function from data on the reaction $ep \to e'nX$.
The corresponding cross section is well 
known \cite{pump} and reads
\beq
\left[\frac{dF_2^{p\to n}(x,Q^2)}{dp_T^2\ dz}
\right]_{1a}=
\frac{2g_{\pi}^2}{16\pi^2}\ 
\frac{|t|}{(m_{\pi}^2-t)^2}\ G_1^2(t)\ 
(1-z)^{1+2\alpha'_{\pi}|t|}\ 
F_2^{\pi}(x_{\pi},Q^2).
\label{1}
\eeq

We use the following notations:
$\vec p_T$ is the transverse component of the neutron momentum relative
to the direction of the virtual photon; $z$ is the Lorenz-invariant
ratio of the neutron to proton light-cone momenta. 
The semi-inclusive structure function $F_2^{p\to n}(x,Q^2)$ is related to the
cross section of electroproduction of the neutron,
\beq
\left[\frac{d\sigma(ep\to e'nX)}{dp_T^2\ dz\ dx\ dQ^2}
\right]_{1a}=
\frac{2\pi\alpha_{em}^2}{xQ^4}
\left (2-2y+y^2\right )\ 
\frac{dF_2^{p\to n}(x,Q^2)}{dp_T^2\ dz}\ ,
\label{2}
\eeq
where $y=Q^2/xs$; $s$ is the c.m. $\gamma^*p$ 
total energy squared. We neglect the small contribution
to the cross section of the longitudinally polarized photons.
The Bjorken variable for the pion is
\beq
x_{\pi}=\frac{x_p}{1-z-x_p}\ ,
\label{4}
\eeq
The $\pi^0NN$ coupling $g_{\pi}$ is fixed at $g_{\pi}^2/4\pi = 13.75$
\cite{g-pi,as}.

The pion four-momentum squared is related to the observables as
\beq
t=-{1\over z} \left [p_T^2 + (1-z)^2m_N^2 \right ]\ .
\label{3}
\eeq

The common $t$-dependence of the formfactor for the pion-nucleon vertex 
and the virtual pion photoabsorption cross section
is effectively parameterized in the form,
 $G_1(t) = \exp[R_1^2(t-m_{\pi}^2)]$, where the slope $R_1^2$ is to be
fitted to experimental data on reactions dominated by pion exchange. 
Unfortunately the results of such an analysis \cite{ponomarev,ab,abk,mpi,fks}
are not stable, and $R_1^2$
varies from zero (see in \cite{desy}) up to $2\ GeV^{-2}$ \cite{ponomarev}.
We fix $R_1^2 = 0.3\ GeV^{-2}$ according to the results of \cite{ab,abk,mpi}
which seem to us most reliable.

It should be pointed out that the formfactor used in \cite{juelich} 
has a different form, namely $G_1 =
exp[R^2\ t/(1-z)]$. This formfactor was introduced ad hoc in \cite{zoller}
and has no justification experimentally (compare the descriptions of
the data on $\Delta$ production at small $(1-z)$ 
in refs. \cite{zoller} and \cite{ag}), and gives energy-dependent $R_1^2$,
which contradicts the results of the approach based on Feynman diagrams,
leading to an abnormal growth of the effective radius at $z \to 1$.

We take into account the Reggeization of the pion exchange in 
eq.~(\ref{1}), this is why we have
the term $2\alpha'_{\pi}t$ as a power of $(1-z)$, where $\alpha'_{\pi} \approx
1\ GeV^{-2}$ is the slope of the pion Regge trajectory. As this is different
from the statement in \cite{juelich}, the Reggeization {\it is}
important in the kinematical region under discussion. The term
$2\alpha_{\pi}'$ in the effective slope of $p_T^2$-distribution
contributes more than the formfactor radius $R_1^2$.

In the experiment $F_2^{\pi}(x,Q^2)$ will be determined by a fit to
experimental data on electroproduction of the neutron.
In order to obtain a numerical estimate of the cross sections 
we assume $x_{\pi} \ll 1$. Thinking in terms of the constituent quark 
model, one may expect a
simple relation $F_2^{\pi} = 2/3 F_2^p$.
The same relation is also predicted by the low-order perturbative QCD 
calculations \cite{gs}. We use for $F_2^p(x,Q^2)$ the results of the 
fit \cite{kp} to low-$x$
data. We checked that another parameterization from \cite{h1} does not
produce any visible distinction.

\begin{figure}[tbh]
\includegraphics{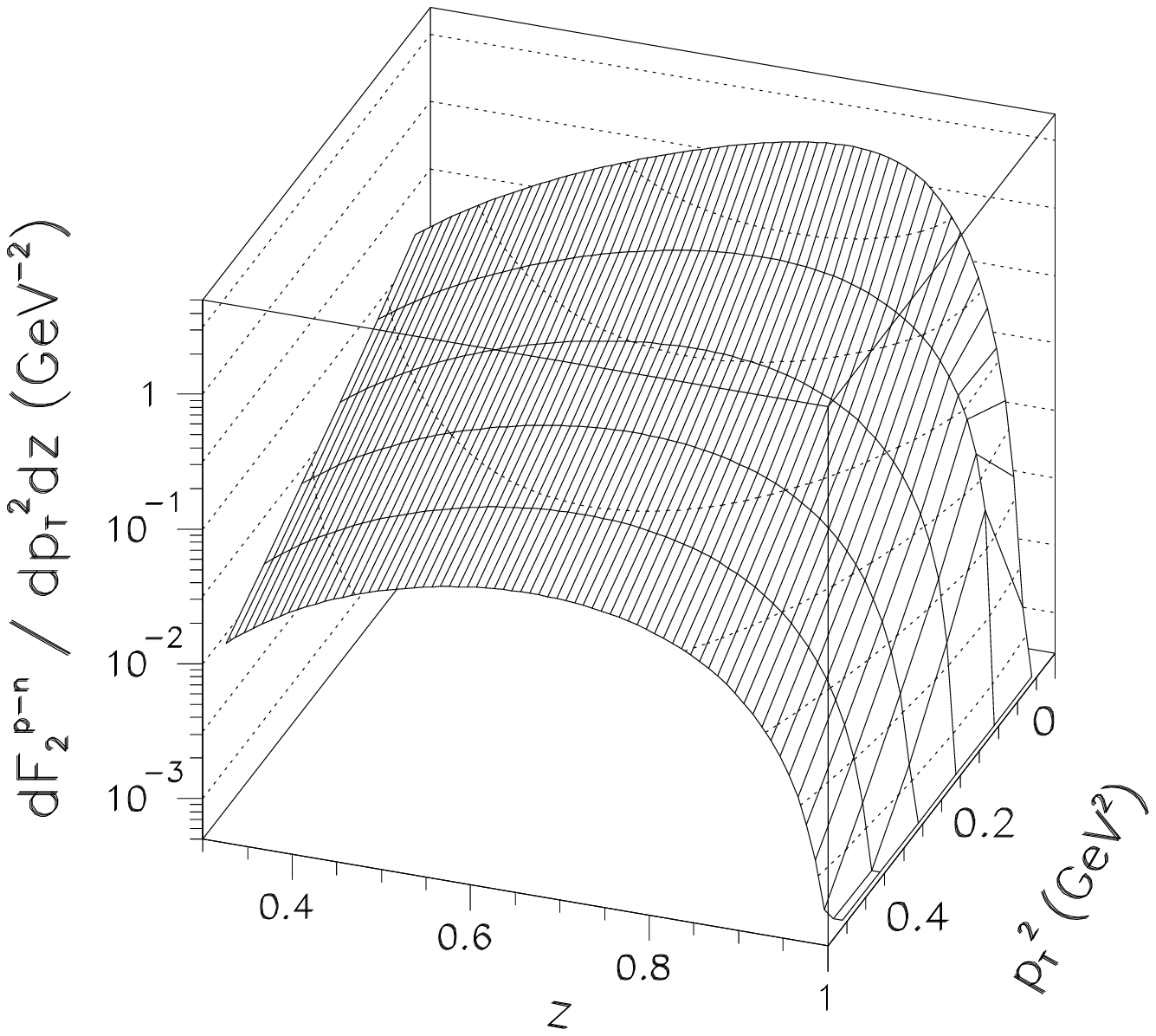}
\begin{center}
\vspace{10cm}
\parbox{13cm}
{\caption[Delta]
{The neutron electroproduction cross section, 
corresponding to
the pion-pole diagram in fig.~\ref{fig1}a,
versus
$p_T^2$ and $z$}
\label{fig3}}
\end{center}
\end{figure}

Our predictions for $z$ - and $p_T^2$-dependence of the 
cross section of inclusive electroproduction of neutrons 
at $Q^2=10\ GeV^2$ and $x_p=5\cdot 10^{-4}$ is depicted
in fig.~\ref{fig3}.

We see that the cross section steeply vanishes towards $z=1$. This is due
to the factor $(1-z)$ in eq.~(\ref{1}), which originates from the
spinelessness of the pion according to the discussion in the previous
section. Small $(1-z)$ correspond to large rapidity gaps, which 
provide a strong suppression of any Regge exchanges with low intercepts.

\section{The pion-pole contribution to $ep \to e'n\pi X$ (fig. 1b)}

As a result of one-pion exchange the proton can be excited, what ends
up with production of extra particles. This contribution was never evaluated,
except for the production of the $\Delta(1236)$ resonance.
The simplest and probably most 
important case is the production of an additional 
pion, as is shown in fig.~\ref{fig1}b.
In this case we have to replace the $\pi NN$ vertex by the $\pi p \to
\pi n$ charge-exchange scattering amplitude. This amplitude, if it is a
realistic one, takes into account also the excitation of all resonances
decaying to the $\pi n$ channel.

The neutron electroproduction cross section, corresponding to the diagram
in fig.~\ref{fig1}b, reads

\beqn
\left[\frac{dF_2^{p\to n}(x,Q^2)}{dp_T^2\ dz}
\right]_{1b}=
& = &
\frac{1}{\pi^3}\ 
\int_{(m_N+m_{\pi})^2}^{s_m}ds_1\ s_1\ 
\int_{z}^{z_{m}}dz_1\ \frac{1-z_1}{z_1^2}\ \times
\nonumber \\
& &
F_2^{\pi}(x_{\pi},Q^2)\ 
\frac{d\sigma^{cex}_{\pi p}(s_1,\theta)}{d\Omega}\ 
\int_0^{2\pi} d\Delta \phi\ \frac{G_2^2(t)}{(m_{\pi}^2-t)^2}\ .
\label{6}
\eeqn

Here $M_X$ is, as before, the effective mass of the jet produced in
the $\gamma^*\pi$ interaction. The effective mass squared 
of the $\pi n$ in the
final state is $s_1$. It ranges from the threshold, $(m_N+m_{\pi})^2$, up to
a maximum value $s_m$ which we fix at $s_1 = 5\ GeV^2$. This choice is
dictated by the energy range where reliable data on pion-nucleon 
scattering exist, and is justified by the fast convergence of the
integral over $s_1$.

The relative share of the initial proton light-cone momentum
carried by the final $\pi n$ system, $z_1$, ranges from $z$
up to $z_m$, which is

\beq
z_m = \min \left \{ \frac{zs_1}{m_N^2}\ ;\ 1 \right \}
\label{7}
\eeq

The value of $z_1$ controls the c.m. energy of the
photon-pion collision, $M_X^2 = s(1-z_1)$.

$d\sigma^{cex}_{\pi p}(s_1,\theta)/d\Omega$ is the differential
cross section of the pion-proton charge-exchange scattering, where
$\Omega(\theta,\phi)$ is the c.m. 
solid scattering angle. The polar angle is related to other variables as

\beq
\cos \theta = 
1-\frac{2s_1}{s_1-m_N^2}\ \left(1-{z\over z_1} \right)\ .
\label{8}
\eeq

The four-momentum squared of the virtual pion in this case reads

\beq
-t = \frac {1}{z_1}[(1-z_1)(2-z_1)(s_1-m_N^2) +
(1-z_1)^2 m_N^2 + q_T^2]\ ,
\label{9}
\eeq
where $\vec q_T$ is the transverse momentum transfer by the pion
exchange,

\beq
q_T^2 = p_T^2 + k_T^2 - 2p_Tk_T\cos(\Delta\phi)
\label{10}
\eeq
and $\vec k_T$ is the transverse momentum of the neutron 
relative to the total momentum of the $\pi n$ system,

\beq
k_T = \frac{s_1-m_N^2}{2\sqrt{s_1}}\ \sin\theta\ .
\label{11}
\eeq

The angle $\Delta\phi$ is the azimuthal angles between $\vec q_T$ and $\vec k_T$
of $\gamma^*p \to X(\pi n)$ and $\pi p \to \pi n$ scatterings.

The formfactor $F_2(t) = \exp[R_2^2(t-m_{\pi}^2)]$ effectively 
takes into account 
the dependence of the $\pi p$ charge-exchange amplitude  and the 
virtual pion photoabsorption cross section on
the off-shellness of the pion. The parameter $R_2^2 \approx 0.6\ GeV^{-2}$
 was fitted in \cite{abk,ag} to describe the data on pion production
in $NN$ and $\pi N$ interactions.

The charge-exchange amplitudes, $\pi^0p \to \pi^+n$ and 
$\pi^-p \to \pi^0n$ (both are included in eq.~(\ref{6}))
were calculated using the code generating the spin- and 
isospin-amplitudes for pion-nucleon scattering, which was kindly
rendered to us by Prof. R.A.~Arndt. The code is based on the 
results of recent phase-shift
analyses \cite{as} of available experimental data on elastic and
charge-exchange pion-nucleon scattering. We used $s_m = 5\ GeV^2$ as 
an upper limit of the integration over $s_1$ in eq.~(\ref{6}). 
Most of the data used in the analyses \cite{as} are within this
energy range, and we checked that the cross section eq.~(\ref{6})
well saturates at this value of $s_m$.

\begin{figure}[tbh]
\includegraphics{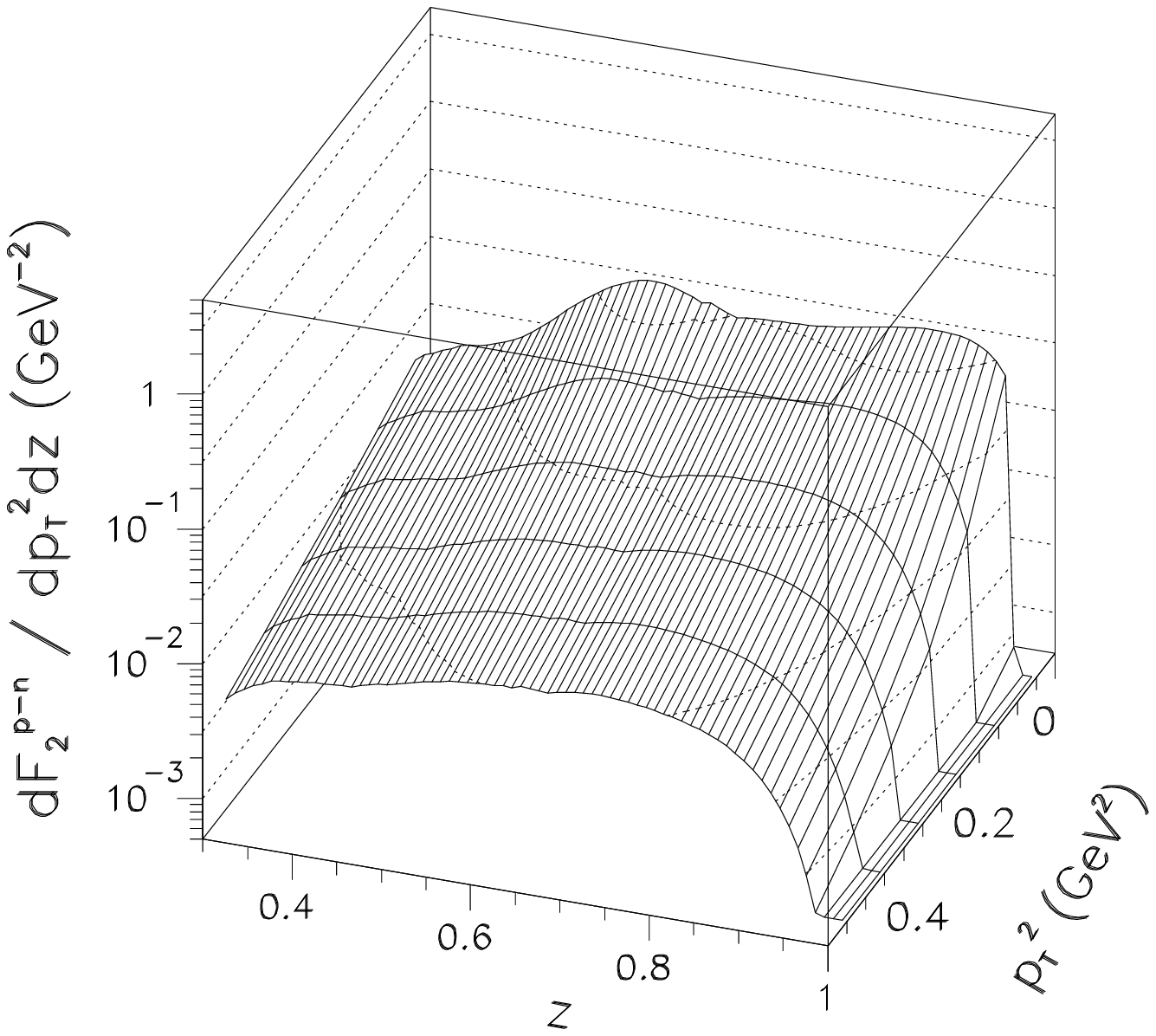}
\begin{center}
\vspace{10cm}
\parbox{13cm}
{\caption[Delta]
{The neutron electroproduction cross section, 
corresponding to
the pion-pole diagram in fig.~\ref{fig1}b,
versus
$p_T^2$ and $z$.}
\label{fig4}}
\end{center}
\end{figure}

The results for the contribution of the diagram in
fig.~\ref{fig1}b are depicted in fig.~\ref{fig4}.
This contribution to the  cross section of electroproduction of the
neutron turns out to be about an order of magnitude smaller than that
for the diagram in fig.~\ref{fig1}a. However, the latter steeply decreases
with the transverse momentum of the neutron, and the role of this
background increases. We study this problem in more detail below.

Note that the $\pi N$ charge-exchange amplitude
presented in the diagram in fig. \ref{1}b may be treated
as a result of $\rho$-Reggeon exchange, provided that 
the c.m. energy squared $s_1$ is sufficiently high. 
This is not the case, since we restricted $s_1 \leq 5\ GeV^2$,
but for higher values of $s_1$ we refer to the next section.

\section{The Reggeon-exchange contribution (fig. 2a)}

As was explained in the Section 2, in order to 
bring the pion closer to its mass shell, one has to increase
the rapidity gap, however, one faces the problem of the increasing 
background
from the leading Reggeon exchange. Those are $\rho$ and $a_2$
Reggeons as is demonstrated in fig.~\ref{fig2}a. 
In this section we make an estimate of this contribution.

The contribution of the mechanism of neutron production shown in
fig.~\ref{fig2}a can be represented in the form
\beq
\left[\frac{dF_2^{p\to n}(x,Q^2)}{dp_T^2\ dz}
\right]_{2a}=
\frac{2g}{16\pi^2}\ 
\left [g_{\rho}^2\ G^2_{\rho}(t) + g_{\omega}^2\ G^2_{\omega}(t) \right ]
(1-z)^{2\alpha'_{R}|t|}\ |\eta_R|^2  
F_2^R(x_R,Q^2)\ ,
\label{12}
\eeq
where $\alpha'_R \approx 0.9\ GeV^{-2}$ is the universal slope of 
the Regge trajectories.

We see that that the $(1-z)$-dependence of
eq.~(\ref{12}) is less steep than that of the pion pole contribution
given by eq.~(\ref{6}). This is because $\rho$ and $a_2$ Regge 
trajectories have
higher intercepts than the pion, $\alpha_{\rho}(0) \approx \alpha_{a_2}(0)
\approx 1/2$.

We neglect the $t$-dependence of the signature factor $\eta_R(t)$ and fix
it at $t=0$, $\eta_{\rho,a_2}(0) = \mp i +1$.

\begin{figure}[tbh]
\includegraphics{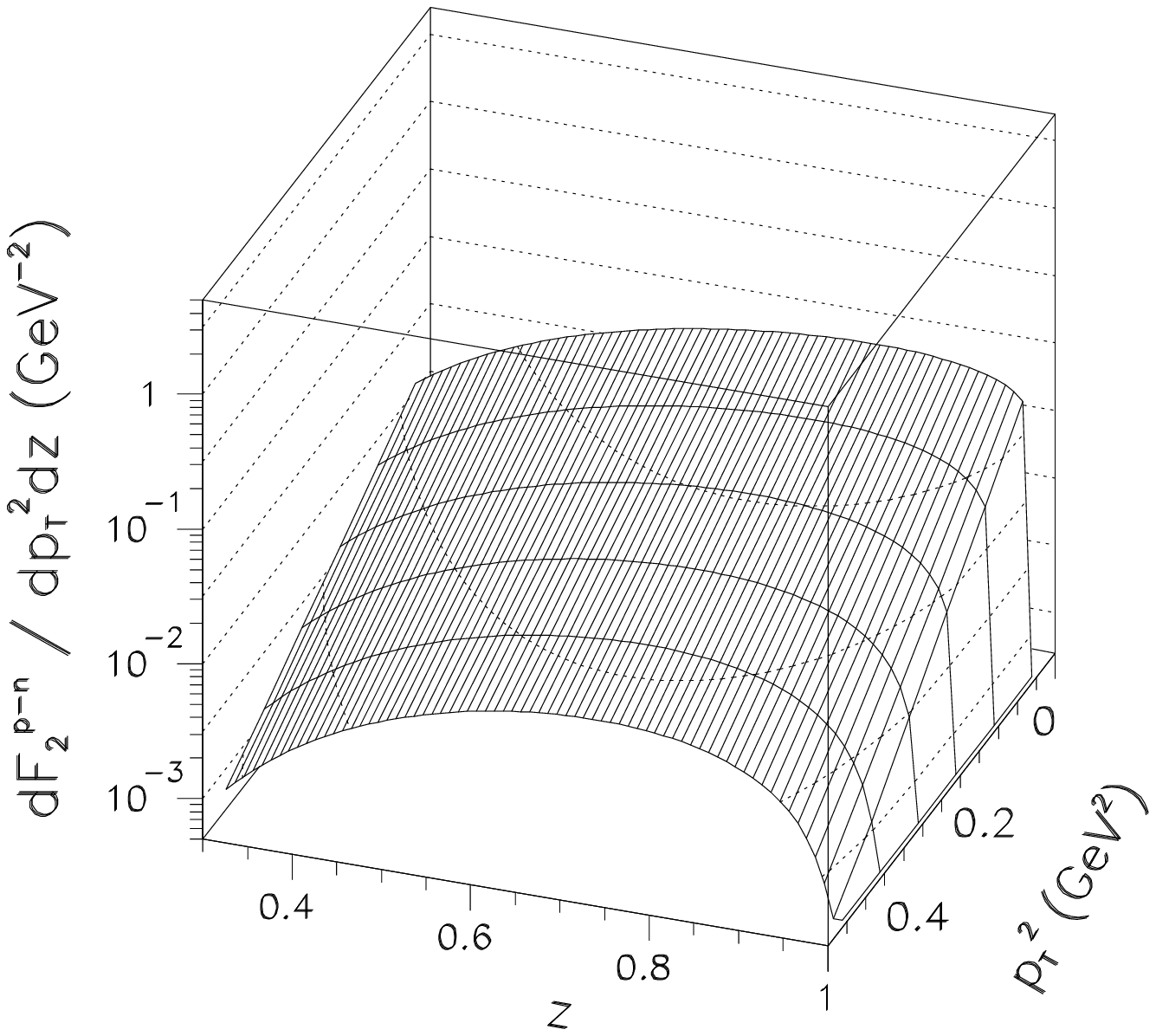}
\begin{center}
\vspace{10cm}
\parbox{13cm}
{\caption[Delta]
{The neutron electroproduction cross section, 
corresponding to
the Reggeon-exchange diagram in fig.~\ref{fig2}a,
versus
$p_T^2$ and $z$.}
\label{fig5}}
\end{center}
\end{figure}

We use the couplings $g_R$ and the formfactors $G_R(t) = \exp(R^2_R\ t)$ 
as they were fixed by the Regge fit \cite{petya} to available experimental
data on high-energy hadronic interactions, $g^2_{\rho}/4\pi = 0.18\ GeV^{-2},\ 
g^2_{a_2}/4\pi = 0.4\ GeV^{-2},\ R^2_{\rho} = 2\ GeV^{-2},\ 
R^2_{a_2} = 1\ GeV^{-2}$. We assume that $F_2^R = F_2^{\pi}$ and $x_R=x_{\pi}$.
The cross section, calculated with eq.~(\ref{12}) and these parameters
is depicted in fig.~\ref{fig5} as a function of $z$ and $t$.

Note that the $a_2$-Reggeon contribution, which is about the same as
from the $\rho$-Reggeon, was missed in ref. \cite{juelich}.

\section{Diffractive electroproduction of neutrons (fig. 2b)}

In the case of double diffractive dissociation the proton may 
produce a jet containing the neutron, for instance like it is shown in 
fig.~\ref{fig2}b. As different from the single-pion exchange in fig.~\ref{fig1}b,
this contribution
is rapidity-gap independent, or can even grow with $\Delta y$ \cite{diff}.
Thus, one may expect a substantial correction from this process.

\begin{figure}[tbh]
\includegraphics{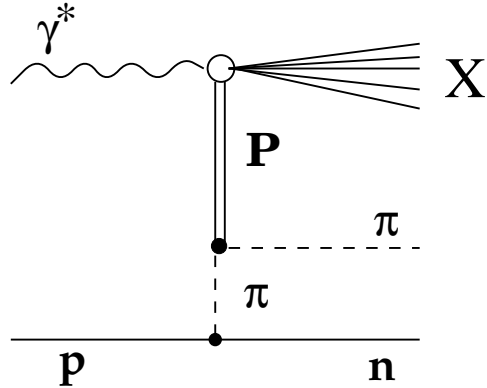}
\begin{center}
\vspace{5cm}
\parbox{13cm}
{\caption[Delta]
{The Deck mechanism of the double diffractive dissociation
electroproduction of the neutron.}
\label{fig6}}
\end{center}
\end{figure}

At $z \to 1$ the reaction in fig.~\ref{fig2}b is 
known to be well described by the so called
Deck mechanism \cite{deck} shown by the diagram in fig.~\ref{fig6}.

This mechanism corresponds to the diffractive dissociation of the photon
on a pion and can be treated as a single-diffraction 
part of the process shown in fig.~\ref{fig1}a. 
The relative contribution of the single photon diffraction to
the total photoabsorption cross section on a pion is expected to be 
about $2/3$ of that in $\gamma^*p$ interaction if one uses the approximate
Pomeron factorization. The photon diffraction on a proton 
was measured 
recently at HERA \cite{diff} at about $10\%$ of the  total photoabsorption
cross section.
Thus, we expect about $6\%$ correction to the measured pion
structure function coming from the diffractive 
mechanism depicted in fig.~\ref{fig6}. On the other hand, even the
proton structure function is known only up to such a correction, so
we neglect it.

\section{Discussion and conclusions}

\begin{figure}[tbh]
\includegraphics{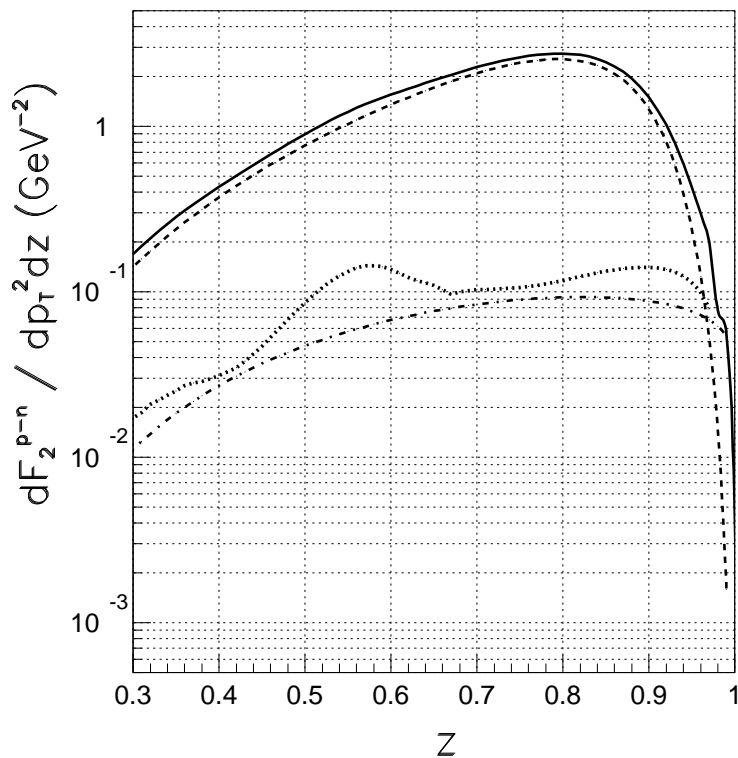}
\begin{center}
\vspace{10cm}
\parbox{13cm}
{\caption[Delta]
{Comparative contributions of the diagrams in fig. 1a
(dashed curve), fig. 1b (dotted curve) 
and fig. 2a (dot-dashed curve) 
to the neutron electroproduction cross section
as a function of $z$ at $p_T^2=0$.
The sum of the three mechanisms is depicted by the solid line.}
\label{fig7}}
\end{center}
\end{figure}

\begin{figure}[tbh]
\includegraphics{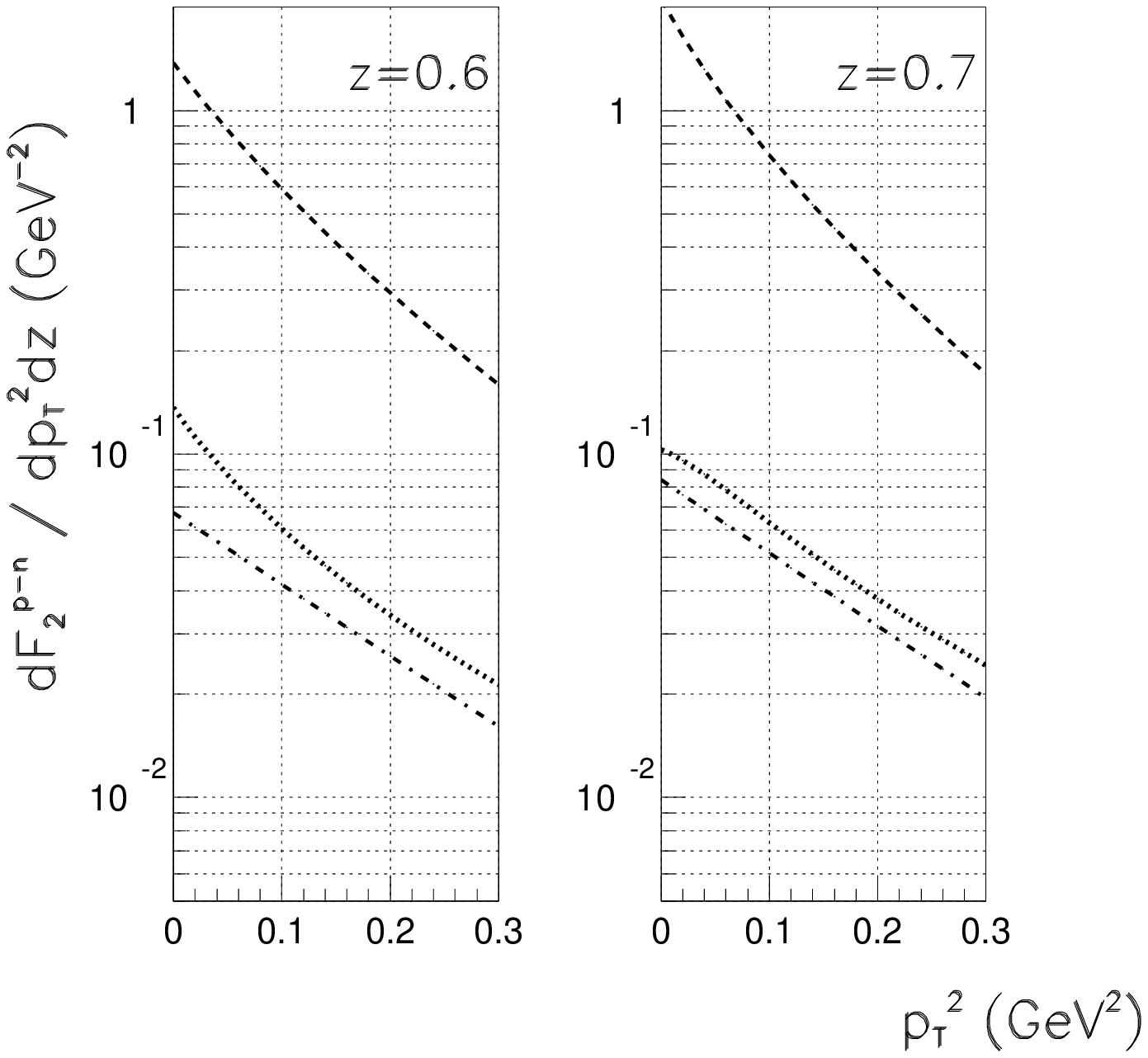}
\includegraphics{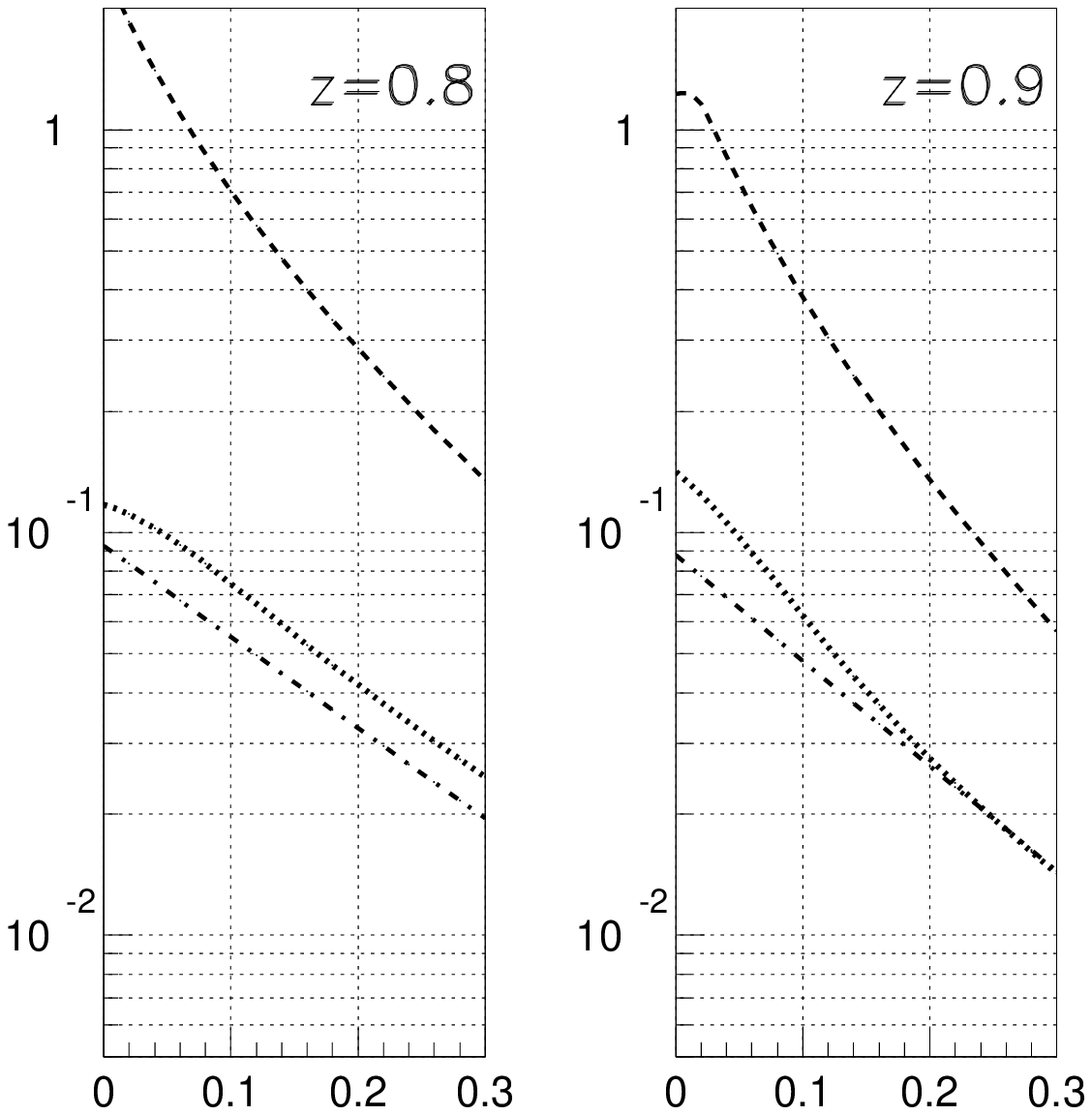}
\begin{center}
\vspace{8cm}
\parbox{13cm}
{\caption[Delta]
{$p_T^2$-distribution of the cross sections of electroproduction
of neutron corresponding to the diagrams in fig.~1a
(dashed curve) fig.~1b (dotted curve) and fig.~2a
(dashed-dotted curve).}
\label{fig8}}
\end{center}
\end{figure}

The contributions of the diagrams in figs.~\ref{fig1}a,b and \ref{fig2}a,
presented in figs.~\ref{fig3}-\ref{fig5}
show that the first one,
fig. \ref{1}a dominates by about an order of magnitude at small 
$p_T^2 \leq 0.3\ GeV^2$. This is demonstrated also on the $z$-distribution
at $p^2_T=0$ in fig.~\ref{fig7} and on $p_T^2$-dependence in fig.~\ref{fig8}.

We should comment, however, on the reliability of our predictions.
In the process under discussion one measures the structure function
of the virtual pion, so a problem arises concerning the 
extrapolation of the result
to the pion pole.
There are quite a few theoretical schemes known in the literature,
which treat the off-mass-shell pion in different ways. 

The approach 
based on the dispersion
relations over the pion four-momentum squared treats the pion pole explicitly,
but one gets in trouble trying to include many
other singularities, $3\pi$-cut, $\pi'$-pole etc.

Using Feynman diagrams one takes into account off-shellness of the pion
by means of the vertex function $G_1(t)$ in eq.~(\ref{1}).
This formfactor is supposed to incorporate effectively 
other singularities in the dispersion relation.
However, an additional $z$-dependence of the formfactor 
is possible because some of the 
further singularities in the dispersion relation correspond to
particles of higher spin, which results in a different energy-,
or $z$-dependence ($1-z=M_X^2/2m_N\nu$) 
of the interaction amplitude compared to that for the pion pole. Usually the
modification of $z$-dependence is neglected in the approach based on 
Feynman diagrams \cite{sull}.

In order to incorporate with different energy dependences corresponding to
the exchange of states with different spins the Regge scheme was suggested
many years ago, which turned out to be 
extremely successful describing the energy 
dependence for many hadronic reactions.
The main idea is to use a more appropriate variable, rather than $t$, 
for the dispersion approach,
which is the complex angular
momentum.  Following \cite{pump,abk} we use the Reggeized
pion exchange. Our eq.~(\ref{1}) contains an additional $z$-dependent
factor $\exp[-2\alpha'_{\pi}t\ln(1-z)]$ compared with the Feynman diagram.
This factor is of special importance at small $(1-z)$ or moderately large
$t$.
The wide-spread opinion that Reggeization is not important for the pion,
since $\alpha_{\pi}(0) \approx 0$, is true only at $t=0$.

A $z$-dependence of the cross section, 
different from the standard Regge approach was suggested in 
\cite{juelich} (and references therein). 
It is based on the light-cone decomposition of
the nucleon over the Fock components containing extra pions.
Although the factorization is not proved for such a decomposition,
it might be a reasonable approximation. Neglecting the components
with two or more pions one comes to the conclusion \cite{zoller} 
that a natural variable for the formfactor for the off-shell pion
is the effective mass squared of the pion-nucleon fluctuation,
which is equivalent to $t/(1-z)$. This results in a
$z$-dependence very different from 
the one dictated by Regge theory, and in an unusual
$t$-dependence which becomes very steep towards $z=1$.
Although it was claimed in \cite{juelich} that such an approach
cannot be used at high $z$ where Reggeization of the pion
is important, the main results of \cite{juelich} 
are obtained in the kinematical region $z > 0.7$, which
is known to be a domain of triple-Regge phenomenology. What is especially
important, theoretical predictions are reliable only at $z > 0.7$ 
(see below), where the approach used in \cite{juelich} completely fails.

The $\rho$-meson exchange is treated in \cite{juelich} on  the
same footing. The failure of such an approach can be easily demonstrated
on a $\rho$-dominated reaction, for instance, $\pi^- p \to \pi^0 n$, which
is well known to be governed by the $\rho$-Reggeon exchange.
Besides, the $a_2$-exchange, neglected in \cite{juelich} is as important
in the diagram in fig.~\ref{fig2}a, as the $\rho$, due to exchange degeneracy.

The distribution over $z$ can be reliably predicted 
only at high $z \geq 0.7$. Indeed, the $z$-dependence of the 
contribution of the diagram in fig.~\ref{fig1}a is controlled by the
energy dependence corresponding to the Reggeized pion exchange, which
we know quite well. However, the left-side slope of the
peak in $z$-distribution in fig.~\ref{fig7} is only due to
the growth of $t$ with decreasing $z$ in accordance with eq.~(\ref{3}).
As a result, the cross section falls down due to the pion
propagator  and the formfactor $G_1^2(t)$ in eq.~(\ref{1}).
The latter, as was commented above, is poorly known, and the results
presented in fig.~\ref{fig7} correspond to an optimistic case of small 
radius $R_1^2 = 0.3/ GeV^{-2}$. Nevertheless, even with a large
radius $R_1^2 = 2\ GeV^{-2}$, the pion pole contribution
dominates by an order of magnitude at $z \approx 0.7-0.8$. 
To be safe one should study the pion-pole
contribution in this kinematical region.

The $p_T^2$-dependence may be substantially
affected by the uncertainty in the slope of the formfactor $G_1(t)$
mentioned above.
The results corresponding to the choice of a small radius are 
depicted in fig.~\ref{fig8} and demonstrate the dominance 
of the pion pole. However, at large radius $R_1^2 = 2\ GeV^{-2}$
the relative contribution of the pion pole fig.~\ref{fig1}a
steeply decreases with $p_T^2$. To be safe one should restrict
$p_T^2 < 0.2-0.3\ GeV^2$.

One can conclude from present analyses that the pion-pole contribution
to the deep-inelastic cross section of the leading neutron production
can be reliably determined, and the pion structure function at small $x_{\pi}$
can be well measured. However, 
in order to disentangle the pion
pole contribution (fig.~\ref{fig1}a) and the background one
should select the events around the maximum  
in $z$-distribution shown in fig.~\ref{fig7}. 
Dependent on the statistics,  one can choose an interval
$z \approx 0.7 - 0.9$  and $p_T^2 < 0.2 - 0.3$.

Summarising, the important signature of the contribution 
of the pion pole graph in fig.~\ref{fig1}a is the steep decrease
of the cross section towards $z=1$ at $z > 0.8$ as is shown in 
fig.~\ref{fig7}. This is mainly due to the spinelessness of the pion,
and our predictions in this kinematical region are least
affected by theoretical uncertainties, particularly those 
in the pion-nucleon formfactor.

Another signature of the pion-pole, the steep $p_T^2$ 
dependence of the cross section, can be smeared due to the smallness
of the radius in the pion-nucleon formfactor. 
In the $z$-interval of the dominance of the pion-pole
demonstrated in fig.~\ref{fig7} the longitudinal momentum transfer is quite 
small, and the exchanged pion is not far from the mass shell. For instance,
at $z=0.9$ the minimal pion virtuality is $|t|_{min} \approx m^2_{\pi}/2$. This
explains why the slope of the $p_T^2$-distribution depicted in fig.~\ref{fig8}
is so large, $B \approx 10\ GeV^{-2}$, in spite of the smallness of the used
radius, $R_1^2 = 0.3\ GeV^{-2}$. The most part of it, about $7\ GeV^{-2}$, originates from
the pion propagator in eq.~(\ref{1}). The rest comes from the $\pi NN$
formfactor and from the Reggeization of the pion pole. The latter part is about 
the same for the $\rho$- and $a_2$-Reggeons, but the former contribution is much smaller. 

An important conclusion is that, regardless of the theoretical uncertainty, the pion pole
contribution corresponding to the graph in fig.~\ref{fig1}a well dominates
by an order of magnitude in the kinematical region $z\approx 0.7-0.9$
and $p_T^2 \leq 0.2 - 0.3\ GeV^2$.

The real challenge to the experimentalists is to identify the pion-pole contribution
by measuring the $p_T^2$ distribution and the $z$-dependence of the 
produced neutrons for $z\to 1$ with the required accuracy. 
Not a simple task with neutrons of several hundred GeV energy.

\medskip

{\bf Acknowledgements}: We are grateful to M.G.~Ryskin, who participated in the 
calculations at the early stage, D.~Jansen and A.~Thomas for the helpful discussions. 
We appreciate very much the courtesy of Prof.
R.A.~Arndt, who provided us with the code generating the amplitudes of pion-nucleon
interaction, based on the phase-shift analyses \cite{as}. We are thankful to 
I.~Strakovsky, who helped us with running the code, for good comments. 
We thank H.~Holtmann and N.N.~Nikolaev for the discussion of the results of
ref.~\cite{juelich}. 
This work was partially supported by INTAS grant 93-0239.

\end{document}